\documentclass[journal]{IEEEtran}

\usepackage{cite}
\usepackage{amsmath,amssymb,amsfonts}
\usepackage{algorithmic}
\usepackage{graphicx}
\usepackage{textcomp}
\usepackage{multirow}
\usepackage{color}
\usepackage{caption}
\usepackage{booktabs}
\usepackage{subfigure}
\usepackage{url}
\usepackage{xcolor}
\usepackage{soul}

\begin{document}
%
\title{A Dynamic Residual Self-Attention Network \\for Lightweight Single Image Super-Resolution}
%
%
%

\author{Karam~Park,~\IEEEmembership{Student Member,~IEEE,}
        Jae~Woong~Soh,~\IEEEmembership{Student Member,~IEEE,}
        and~Nam~Ik~Cho,~\IEEEmembership{Senior Member,~IEEE}
\thanks{K. Park, J. W. Soh, and N. I. Cho are with the Department
of Electrical and Computer Engineering, INMC, Seoul National University, Seoul, 08826 Korea. E-mail: nicho@snu.ac.kr. N. I Cho is also affiliated with the Graduate School of Data Science, Seoul National University.
This work was supported in part by the National Research Foundation of Korea(NRF) grant
funded by the Korea government(MSIT) (2021R1A2C2007220), and in part by Samsung Electronics Co., Ltd.}}%
\maketitle

\begin{abstract}
Deep learning methods have shown outstanding performance in many applications, including single-image super-resolution (SISR).  With residual connection architecture, deeply stacked convolutional neural networks provide a substantial performance boost for SISR, but their huge parameters and computational loads are impractical for real-world applications. Thus, designing lightweight models with acceptable performance is one of the major tasks in current SISR research. The objective of lightweight network design is to balance a computational load and reconstruction performance. Most of the previous methods have manually designed complex and predefined fixed structures, which generally required a large number of experiments and lacked flexibility in the diversity of input image statistics. In this paper, we propose a dynamic residual self-attention network (DRSAN) for lightweight SISR, while focusing on the automated design of residual connections between building blocks. The proposed DRSAN has dynamic residual connections based on dynamic residual attention (DRA), which adaptively changes its structure according to input statistics. Specifically, we propose a dynamic residual module that explicitly models the DRA by finding the interrelation between residual paths and input image statistics, as well as assigning proper weights to each residual path. We also propose a residual self-attention (RSA) module to further boost the performance, which produces 3-dimensional attention maps without additional parameters by cooperating with residual structures. The proposed dynamic scheme, exploiting the combination of DRA and RSA, shows an efficient trade-off between computational complexity and network performance. Experimental results show that the DRSAN performs better than or comparable to existing state-of-the-art lightweight models for SISR.
\end{abstract}


%
\IEEEpeerreviewmaketitle

\section{Introduction}
%
%
%
%

\IEEEPARstart{S}{ingle} image super-resolution (SISR) is one of the image restoration tasks that attempts to find lost information in a low-resolution (LR) image to reconstruct its original high-resolution (HR) counterpart. SISR is a popular research field due to its wide applicability in medical imaging \cite{peled2001superresolution,shi2013cardiac}, satellite imaging \cite{thornton2006sub}, HDTV \cite{goto2014super}, face recognition \cite{gunturk2003eigenface}, surveillance \cite{zhang2010super,rasti2016convolutional}, etc.
However, it is challenging because of its ill-posedness, meaning that there can be multiple possible HR images for a single LR image. 

In recent years, convolutional neural networks (CNNs) have provided considerable performance improvements in computer vision tasks, including SISR. As a pioneer work, Dong \emph{et al.} proposed SRCNN \cite{dong2015image}, which learns a mapping from an interpolated LR image to an HR image. With a simple 3-layer CNN, they showed that the deep learning-based method could outperform the classical SR methods \cite{prev1,prev2,prev3,prev4}.

Meanwhile, researchers have made substantial performance gains for many computer vision tasks by using deep neural networks with residual learning \cite{he2016deep}. Residual learning resolves the deep network training degradation problem, improving network performance and reducing optimization difficulty with simple shortcut connections and an elementwise addition. Due to its simplicity and effectiveness, residual learning has been widely adopted to build deeper and more accurate networks, leading to deep network trends \cite{szegedy2017inception,chollet2017xception}. Influenced by this trend, various deep SR models have been proposed \cite{Kim_2016_VDSR,lim2017enhanced,qiu2019embedded}. However, deepening the network accompanies an increase in parameters and computations. Specifically, while the 3-layer SRCNN \cite{dong2015image} requires only $57\,$K parameters and $52.7\,$G operations, as one of the recent state-of-the-art (SOTA) models with much deeper layers, EDSR \cite{lim2017enhanced}, requires $40.73\,$M parameters and $9384.7\,$G operations. Although deepening the network increases the performance, the accompanying computational load limits real-world applications, such as for a mobile on-device or a real-time SR. For practical applications considering limited computational resources, the demand for lightweight networks has been growing \cite{kim2016deeply, tai2017image, ahn2018fast, chu2019fast, wang2019lightweight, wang2020lightweight}.

The key issue in constructing a lightweight network is balancing the trade-off between the network performance and the cost in terms of the number of parameters (memory complexity) or floating-point operations (time complexity). Since simply reducing the number of layers or blocks of plain networks leads to a severe performance drop, several approaches for designing an efficient network have been suggested. In early studies of lightweight networks, recursive network structures were proposed \cite{kim2016deeply, tai2017image, tai2017memnet}, which reduced the number of parameters by reusing the weights of the network. Although this approach can reduce the number of parameters, it fails to handle the number of operations.

To simultaneously handle the computation and the parameters, networks with efficient and compact design have been proposed \cite{ahn2018fast,choi2018single,wang2019lightweight,hang2020attention,wang2020lightweight,zhao2020efficient,lan2020madnet,tian2020coarse,behjati2021overnet,li2021lightweight,yan2021lightweight}, showing a comparable performance to the SOTA methods. These networks actively exploit shortcut connections to deliver features from the preceding layers to the following layers effectively. With weighted sum methods \cite{wang2019lightweight,wang2020lightweight,behjati2021overnet} or concatenation and pixel-wise convolution methods \cite{ahn2018fast,wang2019lightweight,hang2020attention,wang2020lightweight,lan2020madnet,tian2020coarse,behjati2021overnet}, the delivered features could yield new information efficiently. In addition to feature reuse strategies, the attention mechanism has also been a key tool for maximizing potential performance. For lightweight networks, attention methods with low computational overhead are preferred, such as 1-dimensional (1D) channel attention (CA) \cite{hu2018squeeze} and its variants \cite{wang2020lightweight,li2021lightweight}. Some lightweight networks use more complex attention schemes, utilizing 3D attention \cite{zhao2020efficient} or multiple attention methods \cite{woo2018cbam,hang2020attention,yan2021lightweight} to achieve more powerful feature expression. However, these approaches require exhaustive experiments to find an optimal structure, which is difficult and time-consuming. Meanwhile, there is another issue where input images have diverse statistics, which may require different operations for optimal results. Despite this, the networks mentioned above process all of the input images with the same operations, {\em i.e.}, with a fixed structure, which may yield suboptimal results for some types of inputs.

To solve these problems, we propose a dynamic residual self-attention network (DRSAN) that reconstructs a super-resolved image using variable residual connections. Rather than designing a fixed complex network structure similar to previous works, we make the network learn to control its residual connections according to the input image characteristics. Specifically, one of the main ideas in the proposed method is dynamic residual attention (DRA), which enables the control of residual connections depending on the input image features. Additionally, we introduce residual self-attention (RSA), which adaptively recalibrates features with a 3D attention map. It is important to note that this attention scheme jointly works with the residual connections and does not require additional parameters to build an attention map.

The main contributions of this work are as follows:
\begin{enumerate}
\item We propose a dynamic residual network (DRSAN) enabled by DRA and RSA. The network has a flexible residual structure that changes its residual paths, while considering the input image statistics. The adaptive residual connection helps the network exploit various combinations of residual features to cope with varying input statistics.

\item The proposed RSA produces a 3D attention map working together with residual units. Note that RSA can boost the residual network's performance without an additional module or complex computation.

\item The proposed lightweight DRSAN performs better than or comparable to the SOTA lightweight SR networks, as shown in Table \ref{sotatable}. The proposed network shows an efficient trade-off between reconstruction performance and computational cost.
\end{enumerate}

We presented a preliminary idea for dynamic residual connection in \cite{park2021single}, which showed the DRA's potential for the SR task. The major difference between this paper and our previous work is that we propose a new RSA idea that boosts the performance without an increase in network parameters. Additionally, to combine RSA and DRA, we design a different network architecture from \cite{park2021single}, which requires fewer parameters and obtains better results.
As a result of implementing the RSA and DRA in a new architecture, we can reduce more than $20\%$ of the parameters, while still maintaining reconstruction quality and achieving a state-of-the-art lightweight SR performance.
In addition, we provide a detailed analysis of the experiments to analyze the behavior of DRA and RSA, which we believe contribute to the understanding of our attention mechanisms for SR.


 

\begin{figure*}[!ht]
	\begin{center}
		\includegraphics[width=0.7\linewidth]{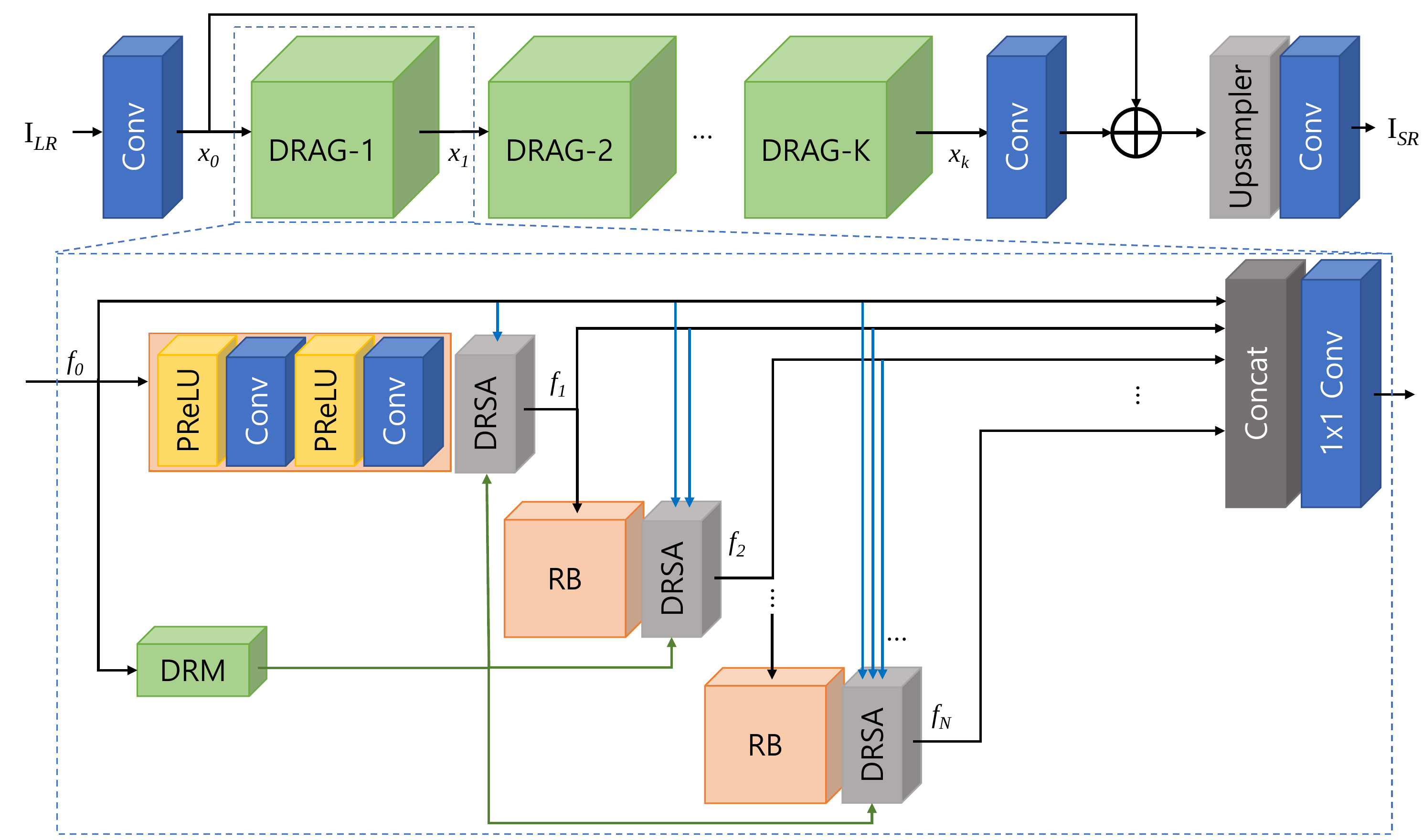}
	\end{center}
	\caption{The framework of the proposed dynamic residual self-attention network (DRSAN). The upper figure shows that it consists of convolution layers (Conv), an upsampling network (Upsampler), and our basic building block DRAGs (dynamic residual attention groups). The lower figure describes the DRAG, which consists of an RB (residual block), a DRSA (dynamic residual self-attention), a DRM (dynamic residual module), a concatenation (Concat), and a $1\times 1$ convolution, where the RB is structured as a cascade of Convs and PReLUs (parametric rectified linear units) \cite{he2015delving}. The signal flow inside the DRAG, including the function of DRSA, is detailed in Fig.~\ref{fig:drsa}.}
	\label{fig:structure}
\end{figure*}

\section{Related Works}
\subsection{CNNs for SISR}

SRCNN \cite{dong2015image} is the first work to apply CNN to the SR task, which outperforms previous works \cite{prev1,prev2,prev3,prev4} with a simple 3-layer CNN.
Dong \emph{et al.} simplified the problem by upsampling the input LR image with bicubic interpolation, turning the problem into refining an upsampled coarse LR image into an HR image. Within this pre-upsampling framework, VDSR \cite{Kim_2016_VDSR} has shown an excellent performance by deepening the network with a residual learning scheme. However, the upsampling strategy causes artifacts, such as amplified noise and blurring. More importantly, the enlarged input image increases the feature size, which is computationally inefficient. ESPCN \cite{shi2016real} introduced an efficient pixel-shuffle layer to reduce the computational inefficiency of previous methods by processing features in the LR space and upsampling it at the end of the network. This post-upsampling strategy has occupied the mainstream of SR research, combined with residual learning. As residual learning \cite{he2016deep} eases training and since post-upsampling improves the computational efficiency, deeper and deeper SISR networks with excellent performances have been proposed \cite{lim2017enhanced,zhang2018image,qiu2019embedded}.
There have also been some methods to propose new architectures and objectives suitable for preserving the structures while super-resolving the images \cite{sr:structure} or for deploying multi-receptive fields with fast computation \cite{MRFN}. For more detailed insights into CNN architectures and objectives for SISR, refer to the review in \cite{yang2019deep}.

\subsection{Lightweight SISR}

Deepening the network guarantees an excellent performance boost at the cost of heavy parameters and computational resources \cite{lim2017enhanced, zhang2018image, zhang2018residual, qiu2019embedded}. Some of the recent models' computational costs exceed certain extents, proving that it is difficult to apply them to real-time scenarios. Hence, various lightweight SR networks have also been proposed for practical applications. For example, DRCN \cite{kim2016deeply}, DRRN \cite{tai2017image}, and DRFN \cite{yang2018drfn} exploited recursive structures to reduce the number of parameters. CARN \cite{ahn2018fast} also showed an acceptable performance with a few parameters using a recursive cascading mechanism. FALSR \cite{chu2019fast} automated the lightweight SR model design based on neural architecture search (NAS) approach \cite{zoph2016neural}. AWSRN \cite{wang2019lightweight} and A$^2$FNet \cite{wang2020lightweight} applied adaptive weights to residual paths, achieving a good balance between the reconstruction performance and the computational cost. While lightweight networks are efficient and effective, most of them are designed manually, requiring much trial and error to achieve an optimal performance. Additionally, these networks have predefined fixed structures, and thus, process all of the images with the same operation, which may not be optimal for accurate SR image reconstruction with limited network capacity. For instance, some features might be informative for a certain input LR image, though it may be redundant for others. In comparison, our method learns the correlation between input image statistics and residual paths during training, and then changes its residual connections depending on the input image. Hence, our method can represent various features as a linear combination of features from preceding blocks.

\subsection{Attention Mechanisms for SISR}
The attention mechanism inspired by neuroscience \cite{mnih2014recurrent}, has brought substantial improvements to neural network applications. By adjusting the weights, the attention mechanism assigns priorities to important features. SENet \cite{hu2018squeeze} proposed a CA scheme, which has been widely adopted in SR networks due to its simplicity, effectiveness, and small computational overheads. The CA gives a 1D (1$\times$1$\times$C) attention map for scaling the features by channel-wise importance. Roy \emph{et al.} \cite{roy2018concurrent} proposed spatial attention (SA), obtained from spatial-wise squeeze and excitation. The SA yields a 2D (H$\times$W$\times$1) attention map and prioritizes notable areas. A-CubeNet \cite{hang2020attention}, LAMRN \cite{yan2021lightweight}, GLADSR \cite{zhang2020accurate}, and MAMSR \cite{yang2021image} jointly utilize SA and CA, showing that using multiple attention methods can improve the reconstruction performance. SAN \cite{dai2019second} proposed another attention mechanism and considered second-order feature statistics for accurate SR reconstruction. Recently, inspired by SCNet \cite{liu2020improving}, Zhao \emph{et al.} \cite{zhao2020efficient} proposed a pixel attention (PA) scheme for a lightweight network. The network computes a 3D attention map with a small computational cost, showing that this type of map is effective for accurate SR reconstruction. While these previous works have shown excellent performances, they have only focused on the features from the residual blocks and did not consider the features from the skip connections. In addition, the existing methods have been gradually improved by using increasingly complex attention methods, which require additional attention modules and higher computational costs. Compared to previous methods, we apply the attention scheme to skip connections, named DRA, which allows the network to utilize various combinations of features from preceding layers to cope with the varying input statistics. This dynamic structure gives the network flexibility against diverse image statistics, boosting the network's representational power with a limited capacity. Additionally, we propose RSA, which generates a 3D attention map with the features from the residual block. By cooperating with a residual structure, RSA can generate attention coefficients without additional modules or complex computation.

\section{Proposed Method}
\subsection{Network Structure}

As shown in Fig.~{\ref{fig:structure}}, the proposed DRSAN consists of three modules: a feature extraction module (Conv), a non-linear mapping module with multiple dynamic residual attention groups (DRAGs), and a reconstruction module (Upsampler and Conv).
\subsubsection{Feature Extraction Module}
The feature extraction module is the first convolutional layer with a kernel size of 3. The input LR image is fed to the first convolutional layer, which can be formulated as follows:
\begin{equation}x_0=\text{F}_{ext}(\text{I}_{LR}),\label{featureextraction}\end{equation}
where $\text{F}_{ext}$ denotes the feature extraction operation and $\text{I}_{LR}$ denotes the LR input image. The first convolutional layer performs shallow feature extraction, and the output feature $x_0$ is sent to the non-linear mapping module and the reconstruction module.

\subsubsection{Nonlinear Mapping Module}
The nonlinear mapping module consists of multiple DRAGs, which extract information to reconstruct an HR image. The shallow feature $x_0$ is processed with stacked DRAGs as follows:
\begin{equation}x_k=\text{F}_{DRAG}^k(x_{k-1}),\; k=1,2,...,K,\label{nonlinearmapping}\end{equation}
where $\text{F}_{DRAG}^k$ denotes the DRAG operation of the $k$-th DRAG, and $K$ denotes the number of DRAGs in the DRSAN. The proposed DRAG consists of multiple residual blocks (RBs), dynamic residual module (DRM), and dynamic residual self-attention (DRSA) modules, as illustrated in the lower part of Fig.~{\ref{fig:structure}}. Additionally, the signal flow in a part of the DRAG is described in Fig.~\ref{fig:drsa}. The figure shows that the DRM controls residual connections to adaptively combine the features, allowing the DRAG to combine the features of residual blocks with appropriate weights to deal with various input image properties. The figure also illustrates that DRSA exploits two attention mechanisms: DRA (upper part of the gray box) and RSA (lower part). The details of the DRAG and its attention mechanism are described in the next subsection (Sec. III.B).
\subsubsection{Reconstruction Module}
Extracted deep feature $x_K$ is then upscaled through the upsampler network, which is described as
\begin{equation}x_{up}=\text{G}_{up}(\text{F}_{3\times 3}(x_K)+x_0),\label{reconstruction1}\end{equation}
where $\text{G}_{up}$ denotes the upsampling function, which consists of one convolutional layer and the pixel-shuffle layer \cite{shi2016real}. Following the global residual learning scheme \cite{Kim_2016_VDSR}, shallow feature $x_0$ is added to deep feature $x_K$ before the pixel-shuffle operation. The pixel-shuffle layer transforms the shape of the input feature with the depth-to-space operation. After upscaling, the network reconstructs the super-resolved (SR) image as follows:
\begin{equation}\text{I}_{SR}=\text{F}_{rec}(x_{up}),\label{reconstruction2}\end{equation}
where $\text{I}_{SR}$ is the reconstructed SR image, and $\text{F}_{rec}$ is the operation of the last convolutional layer with a kernel size of 3.

\subsection{Dynamic Residual Attention Group (DRAG)}
As a building block of the nonlinear mapping module, the DRAG extracts residual features to reconstruct an accurate HR image. It has three main elements: residual blocks, DRMs, and DRSAs.

\subsubsection{Dynamic Residual Module (DRM)}
\begin{figure}[t]
	\begin{center}
		\includegraphics[width=\linewidth]{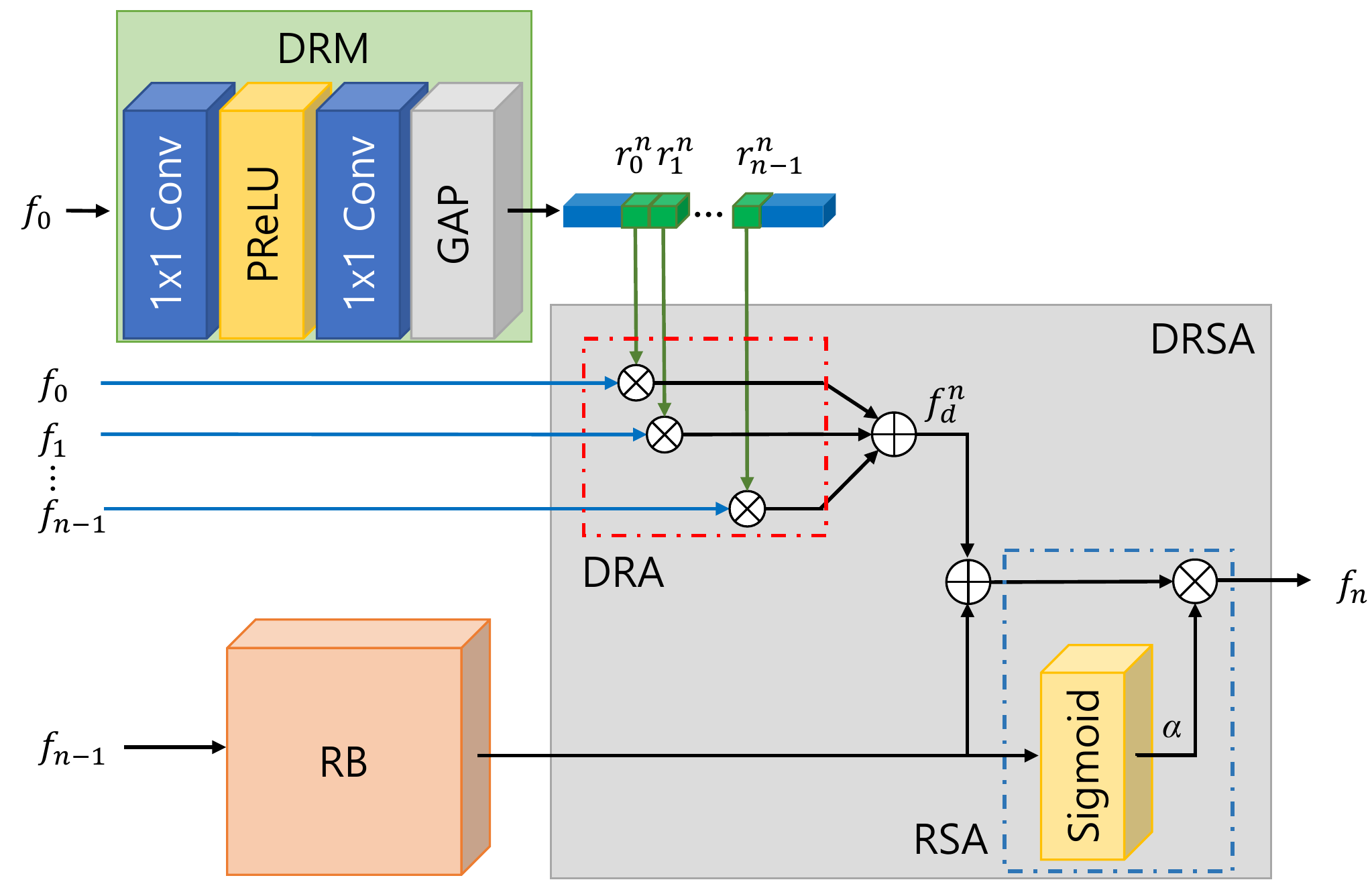}
	\end{center}
	\caption{The signal flow graph inside the DRAG, and the function of the $n$-th DRSA. The DRSA outputs the $n$-th residual feature ($f^n$) as a combination of $f_d^n$ (addition of previous features with DRA) and $\alpha$ (RSA formed by the RB and sigmoid). The DRM determines the DRA that reflects the input properties.}
	\label{fig:drsa}
\end{figure}

Fig.{\ref{fig:drsa}} illustrates the signal flow around and inside the $n$-th RB and DRSA, along with the DRM. The DRM is a simple module with two convolutional layers, which controls residual connections between the residual blocks in a DRAG. Notably, the DRM does not use the sigmoid function as an activation function to exploit residual coefficients in wide ranges for feature diversity.
It decides how to combine the residual features, {\em i.e.,}, computes DRA parameters ($r_k^n$ shown above the DRSA in Fig.~\ref{fig:drsa}), considering the given input feature of a DRAG. Formally, the DRM determines the DRA parameters as follows:
\begin{equation}
\begin{split}
r & =\text{M}_{DRM}(f_0) \\
& = \text{P}_{avg}(\text{F}_{out}(\text{PReLU}(\text{F}_{in}(f_0)))) \\
& = [r_{0}^1,r_{0}^2,r_{1}^2,r_0^3,r_1^3,r_2^3,\cdots,r_{0}^N,r_{1}^N,\cdots,r_{N-1}^N],
\end{split}
\label{dynamicresidualmodule}\end{equation}
where $\text{M}_{DRM}$ denotes DRM operation, $\text{P}_{avg}$ denotes global average pooling, $\text{F}_{in}$ and $\text{F}_{out}$ denote convolutional layers, $f_0$ denotes the input feature of DRAG, and $N$ denotes the number of residual blocks in a DRAG. During the training process, DRM learns the correlation between input image statistics and residual paths, which helps the network exploit proper features among the low- and high-level features with various combinations.

\subsubsection{Dynamic Residual Self-Attention (DRSA)}
The proposed DRSA is an attention scheme that allows the network to exploit various combinations of residual features in DRAGs. As shown in Fig.~{\ref{fig:drsa}} and Eq.~(\ref{dynamicresidualmodule}), DRM determines the DRA parameters $r$ considering the input feature statistics. These parameters scale the corresponding residual paths in DRAG, allowing the network to adaptively change its residual structure for the input image. The residual feature $f_n$ of the $n$-th residual block can be described as

\begin{equation}f_{n}=(\text{F}_{res}^n(f_{n-1})+f_{d}^n) \cdot \alpha, \label{DRAG}\end{equation}
where $\text{F}_{res}^n$ denotes convolution operation of the $n$-th residual block, $f_{d}^n$ denotes dynamic residual feature for the $n$-th residual block, and $\alpha$ denotes the residual self attention coefficient. With the DRA attention parameter $r^n=[r_{0}^n,... ,r_{n-1}^n]$ decided by the DRM, dynamic reisdual feature $f_{d}^n$ can be formulated as

\begin{equation}f_{d}^n=(r_0^n\cdot f_0 + r_1^n\cdot f_1 + ... + r_{n-1}^n\cdot f_{n-1} ). \label{DRA}\end{equation}
This indicates that all features from preceding residual blocks of a DRAG are potentially connected, allowing DRAG to utilize diverse combinations of residual features. After the dynamic residual feature $f_{d}^n$ is computed, the residual self-attention is applied. The residual self-attention coefficient $\alpha$ is calculated as

\begin{equation} \alpha = \sigma(\text{F}_{res}^n(f_{n-1})), \label{eq:RSA}\end{equation}
where $\sigma$ denotes sigmoid activation function. This simple yet efficient attention scheme yields a 3D attention map, which scales features channel-wisely and spatial-wisely. The proposed RSA recalibrates the network features and helps the network focus on more informative features.

After all of the residual features ${f_1,f_2,...,f_N}$ are decided, the output of the $k$-th DRAG $x_k$ can be formulated as follows:

\begin{equation} x_k = \text{F}_{1 \times 1}(\text{concat}[f_0,f_1,...,f_N]),\label{DRAGoutput}\end{equation}
where $\text{F}_{1 \times 1}$ denotes the 1$\times$1 convolution, and $concat$ denotes the channel-wise concatenation of features.


\begin{table}[t]
	\caption{\label{implementationdetail} Implementation details of the proposed method. $N$ means the number of residual blocks in a DRAG, $k$ stands for the $k$-th DRAG, $c$ is the number of channels in a convolutional layer, $s$ is the scaling factor, and GAP indicates the global average pooling.} 
	\renewcommand*{\arraystretch}{1.4}
	\begin{tabular}{|c|c|c|c|c|}
		\hline
		\multicolumn{2}{|c|}{Function}                                        & Layer        & Kernel      & \begin{tabular}[c]{@{}c@{}}Channels\\ (Input, Output)\end{tabular} \\ \hline
		\multicolumn{2}{|c|}{$\text{F}_{ext}$}                                       & Conv         & $3\times 3$ & (3, $c$)                                                            \\ \hline
		\multirow{9}{*}{$\text{F}^k_{DRAG}$} & \multirow{4}{*}{$\text{F}_{res}$}            & PReLU        & -           & -                                                                  \\
		&                                       & Conv         & $3\times 3$ & ($c$, $c$)                                                           \\
		&                                       & PReLU        & -           & -                                                                  \\
		&                                       & Conv         & $3\times 3$ & ($c$, $c$)                                                           \\ \cline{2-5} 
		& \multirow{4}{*}{$\text{M}_{DRM}$}            & Conv         & $1\times 1$ & ($c$, 16)                                                           \\
		&                                       & PReLU        & -           &  -                                                                  \\
		&                                       & Conv         & $1\times 1$ & (16, $\frac{N\cdot(N+1)}{2}$)                                                            \\
		&                                       & GAP         & -           &  -                                                                  \\ \cline{2-5} 
		& $\text{F}_{1\times 1}$                       & Conv         & $1\times 1$ & ($c$$\cdot$N, $c$)                                                         \\ \hline
		\multicolumn{2}{|c|}{$\text{F}_{3\times 3}$}                                 & Conv         & $3\times 3$ & ($c$, $c$)                                                           \\ \hline
		\multirow{6}{*}{$\text{G}_{up}$}       & \multirow{2}{*}{$\times 2, \times 3$} & Conv         & $3\times 3$ & ($c$, $c$$\cdot s^2$)                                                \\
		&                                       & PixelShuffle & -           & -                                                                  \\ \cline{2-5} 
		& \multirow{4}{*}{$\times 4$}           & Conv         & $3\times 3$ & ($c$, $c$$\cdot 2^2$)                                                \\
		&                                       & PixelShuffle & -           & -                                                                  \\
		&                                       & Conv         & $3\times 3$ & ($c$, $c$$\cdot 2^2$)                                                \\
		&                                       & PixelShuffle & -           & -                                                                  \\ \hline
		\multicolumn{2}{|c|}{$\text{F}_{rec}$}                                           & Conv         & $3\times 3$ & ($c$, 3)                                                            \\ \hline
	\end{tabular}
\end{table}


\section{Experimental Results}
\label{sec:guidelines}

\subsection{Dataset}
The DRSAN is trained with DIV2K dataset \cite{agustsson2017ntire}, which consists of 800 high-resolution training images. We use RGB patches with a size of 48$\times$48 for training. Additionally, we perform data augmentation on the training dataset with flips and rotations. We evaluate the peak signal-to-noise ratio (PSNR) and structural similarity index (SSIM) \cite{wang2004image} on the Y channel of the test images, using four SR benchmark datasets: Set5 \cite{bevilacqua2012low}, Set14 \cite{zeyde2010single}, B100 \cite{martin2001database}, and Urban100 \cite{huang2015single}. 

\subsection{Training Details}
The mini-batch size is set to 16. We use ADAM optimizer for network training with $\beta_1 = 0.9$, $\beta_2 = 0.999$, and $\epsilon = 10^{-8}$. The learning rate is initialized as $2\times10^{-4}$, and decayed with the factor of 0.85 for every $2 \times 10^5$ iterations. The L1 is used as a loss function for training.

\subsection{Implementation Details}
To explore the trade-off between depth and channel on our model, we prepare five variants of DRSAN: DRSAN-32s, DRSAN-32m, DRSAN-32l, DRSAN-48s, and DRSAN-48m, which have different numbers of channels and layers. Small-scale models are suffixed by ``-s," medium-scale by ``-m," and large-scale by ``-l," and each model has 4, 8, and 10 DRAGs, respectively. The models using 32 channels for convolutional layers are suffixed by ``-32," and the models using 48 channels by ``-48."
To generate networks with similar complexities to the compared methods, we make the DRAGs have three residual blocks ($N=3$) for DRSAN-48 models and four residual blocks ($N=4$) for DRSAN-32 models.
The DRM in DRAGs has 16 filters for the first convolutional layer.
The second convolutional layer in DRM has 6 filters for 48 channel models and 10 filters for 32 channel models, where 6 and 10  correspond to $\frac{N\cdot(N+1)}{2}$ for $N=3$ and 4, respectively.

As the upsampler module, we use the ESPCN method \cite{shi2016real}. The last convolutional layer in the reconstruction module has three channels to reconstruct an RGB image. More details about the model structure are described in Table~\ref{implementationdetail}. All results are evaluated on the NVIDIA TITAN XP GPU device.

\begin{figure}[!t]
	\begin{center}
		\includegraphics[width=\linewidth]{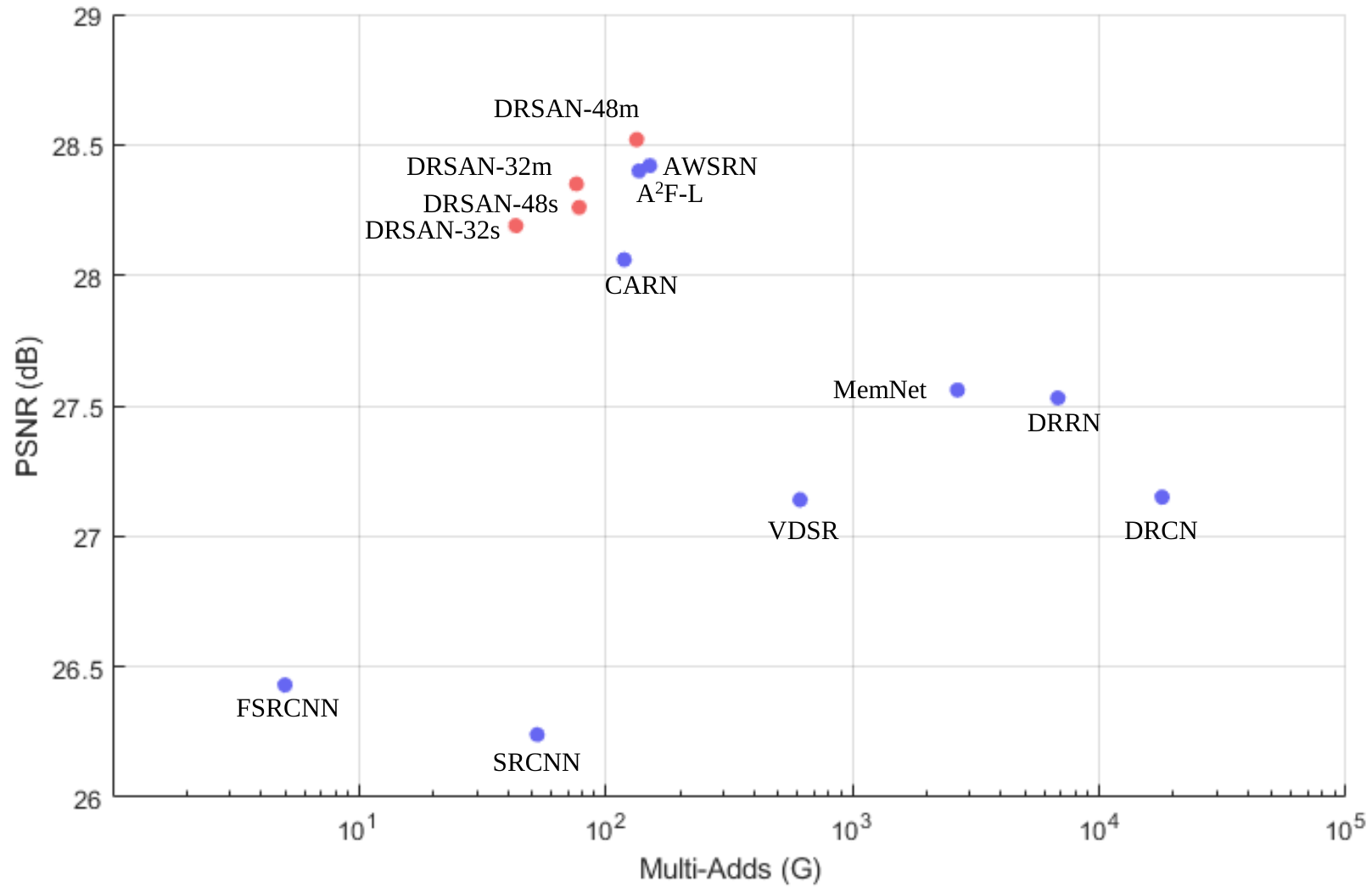}
	\end{center}
	\caption{Performance comparison between the SOTA lightweight networks and our methods. The results are evaluated on the Urban100 ($\times3$) dataset.}
	\label{fig:PvO}
\end{figure}

\begin{table*}
	\caption{\label{sotatable} Quantitative comparisons of our models with the previous SOTA lightweight networks on benchmark datasets. Our models are marked in \textbf{bold}. The best performance is denoted in \textcolor{red}{red}, and the second-best performance is denoted in \textcolor{blue}{blue}. ``Param'' is the number of parameters, and ``Multi-Adds'' is the number of multiply-accumulate operations.} 
	\begin{center}
	    \resizebox{0.85\linewidth}{!}{
		\begin{tabular}{cllclcccc}
			\toprule[1pt]
			\multirow{2}{*}{Scale} & \multirow{2}{*}{Scope}  & \multirow{2}{*}{Model} & \multirow{2}{*}{Param} & \multirow{2}{*}{Multi-Adds} & Set5 & Set14 & B100 & Urban100       \\
			&                         &                        &                        &                 & PSNR / SSIM            & PSNR / SSIM      & PSNR / SSIM      & PSNR / SSIM      \\ \midrule\midrule
			\multirow{27}{*}{$\times 2$} & \multirow{9}{*}{$<$ 500$\,$K} & FSRCNN                 & 0.01$\,$M                  & 6.0G                & 37.00 / 0.9558 & 32.63 / 0.9088 & 31.53 / 0.8920 & 29.88 / 0.9020 \\
			& & DRRN                   & 0.30$\,$M                  & 6796.9G             & 37.74 / 0.9591 & 33.23 / 0.9136 & 32.05 / 0.8973 & 31.23 / 0.9188 \\
			& & A$^2$F-SD              & 0.31$\,$M                  & 71.2G               & \textcolor{blue}{37.91 / 0.9602} & \textcolor{blue}{33.45 / 0.9164} & \textcolor{blue}{31.79 / 0.8986} & \textcolor{blue}{31.79 / 0.9246} \\
			& & A$^2$F-S               & 0.32$\,$M                  & 71.7G               & 37.79 / 0.9597 & 33.32 / 0.9152 & 31.99 / 0.8972 & 31.44 / 0.9211 \\
			& & FALSR-B                & 0.33$\,$M                  & 74.7G               & 37.61 / 0.9585 & 33.29 / 0.9143 & 31.97 / 0.8967 & 31.28 / 0.9191 \\
			& & AWSRN-SD               & 0.35$\,$M                  & 79.6G               & 37.86 / 0.9600 & 33.41 / 0.9161 & 32.07 / 0.8984 & 31.67 / 0.9237 \\
			& & AWSRN-S                & 0.40$\,$M                  & 91.2G               & 37.75 / 0.9596 & 33.31 / 0.9151 & 32.00 / 0.8974 & 31.39 / 0.9207 \\
			& & FALSR-C                & 0.41$\,$M                  & 93.7G               & 37.66 / 0.9586 & 33.26 / 0.9140 & 31.96 / 0.8965 & 31.24 / 0.9187 \\ 
			& & \textbf{DRSAN-32s (Ours)}     & \textbf{0.37$\,$M}         & \textbf{85.5G}      & \textcolor{red}{\textbf{37.99 / 0.9606}} & \textcolor{red}{\textbf{33.57 / 0.9177}} & \textcolor{red}{\textbf{32.16 / 0.8999}} & \textcolor{red}{\textbf{32.10 / 0.9279}} \\ \cmidrule[0.1pt]{2-9}
			& \multirow{10}{*}{$<$ 1000$\,$K} & IDN                   & 0.55$\,$M                  & -              & 37.83 / 0.9600 & 33.30 / 0.9148 & 32.08 / 0.8985 & 31.27 / 0.9196 \\
			& & VDSR                   & 0.66$\,$M                  & 612.6G              & 37.53 / 0.9587 & 33.03 / 0.9124 & 31.90 / 0.8960 & 30.76 / 0.9140 \\
			& & MemNet                 & 0.68$\,$M                  & 2662.4G              & 37.78 / 0.9597 & 33.28 / 0.9143 & 32.08 / 0.8978 & 31.31 / 0.9195 \\
			& & IMDN                   & 0.69$\,$M                  & -              & 38.00 / 0.9605 & 33.63 / 0.9177 & 32.19 / 0.8996 & 32.17 / 0.9283 \\
			& & LapSRN                 & 0.81$\,$M                  & 29.9G              & 37.52 / 0.9590 & 33.08 / 0.9130 & 31.80 / 0.8950 & 30.41 / 0.9100 \\
			& & SelNet                 & 0.97$\,$M                  & 225.7G               & 37.89 / 0.9598 & 33.61 / 0.9160 & 32.08 / 0.8984 & -              \\
			& & A$^2$F-M               & 1.00$\,$M                  & 224.2G               & 38.04 / 0.9607 & \textcolor{blue}{33.67} / 0.9184 & 32.18 / 0.8996 & 32.27 / 0.9294 \\ 
			& & \textbf{DRSAN-48s (Ours)} & \textbf{0.65$\,$M}         & \textbf{150.0G}     & \textbf{38.08 / \textcolor{blue}{0.9609}} & \textbf{33.62 / 0.9175} & \textbf{32.19 / 0.9002} & \textbf{32.16 / 0.9286} \\
			& & \textbf{DRSAN-32m (Ours)}     & \textbf{0.69$\,$M}         & \textbf{159.3G}      & \textcolor{blue}{\textbf{38.11 / 0.9609}} & \textbf{33.64 / \textcolor{blue}{0.9185}} & \textcolor{blue}{\textbf{32.21 / 0.9005}} & \textcolor{blue}{\textbf{32.35 / 0.9304}} \\ 
			& & \textbf{DRSAN-32l (Ours)}     & \textbf{0.85$\,$M}         & \textbf{196.3G}      & \textcolor{red}{\textbf{38.13 / 0.9610}} & \textbf{\textcolor{red}{33.72 / 0.9189}} & \textcolor{red}{\textbf{32.24 / 0.9009}} & \textcolor{red}{\textbf{32.41 / 0.9312}} \\ \cmidrule[0.1pt]{2-9}
			&\multirow{8}{*}{$<$ 2000$\,$K} & FALSR-A                & 1.02$\,$M                  & 234.7G               & 37.82 / 0.9595 & 33.55 / 0.9168 & 31.12 / 0.8987 & 31.93 / 0.9256 \\
			& & AWSRN-M                & 1.06$\,$M                  & 244.1G               & 38.04 / 0.9605 & 33.66 / 0.9181 & 32.21 / 0.9000 & 32.23 / 0.9294 \\
			& & A$^2$F-L               & 1.36$\,$M                  & 306.1G               & 38.09 / 0.9607 & 33.78 / 0.9192 & 32.23 / 0.9002 & 32.46 / 0.9313 \\
			& & OISR-RK2-s             & 1.37$\,$M                  & 316.2G               & 37.98 / 0.9604 & 33.58 / 0.9172 & 32.18 / 0.8996 & 32.09 / 0.9281 \\
			& & AWSRN                  & 1.40$\,$M                  & 320.5G               & \textcolor{blue}{38.11 / 0.9608} & \textcolor{red}{33.78 / 0.9189} & \textcolor{red}{32.26} / \textcolor{blue}{0.9006} & \textcolor{red}{32.49} / \textcolor{blue}{0.9316} \\
			& & CARN                   & 1.59$\,$M                  & 222.8G               & 37.76 / 0.9590 & 33.52 / 0.9166 & 32.09 / 0.8978 & 31.92 / 0.9256 \\
			& & DRCN                   & 1.77$\,$M                  & 17974G               & 37.63 / 0.9588 & 33.04 / 0.9118 & 31.85 / 0.8942 & 30.75 / 0.9133 \\ 
			& & \textbf{DRSAN-48m (Ours)}  & \textbf{1.19$\,$M}         & \textbf{274.6G}      & \textcolor{red}{\textbf{38.14 / 0.9611}} & \textcolor{blue}{\textbf{33.75 / 0.9188}}  & \textcolor{blue}{\textbf{32.25}} \textbf{/} \textcolor{red}{\textbf{0.9010}} & \textcolor{blue}{\textbf{32.46}} \textbf{/} \textcolor{red}{\textbf{0.9317}} \\ \midrule
			\multirow{21}{*}{$\times 3$} & \multirow{7}{*}{$<$ 500$\,$K} & FSRCNN                 & 0.01$\,$M                  & 5.0G                 & 33.16 / 0.9140 & 29.43 / 0.8242 & 28.53 / 0.7910 & 26.43 / 0.8080 \\
			& & DRRN                   & 0.30$\,$M                  & 6797G                & 34.03 / 0.9244  & 29.96 / 0.8349 & 28.95 / 0.8004 & 27.53 / 0.8378 \\
			& & A$^2$F-SD              & 0.32$\,$M                  & 31.9G               & \textcolor{blue}{34.23} / 0.9259 & \textcolor{blue}{30.22} / 0.8395 & \textcolor{blue}{29.01 / 0.8028} & \textcolor{blue}{27.91 / 0.8465} \\
			& & A$^2$F-S               & 0.32$\,$M                  & 32.3G               & 34.06 / 0.9241 & 30.08 / 0.8370 & 28.92 / 0.8006 & 27.57 / 0.8392 \\
			& & AWSRN-SD               & 0.39$\,$M                  & 39.5G                & 34.18 / \textcolor{red}{0.9273} & 30.21 / \textcolor{blue}{0.8398} & 28.99 / 0.8027 & 27.80 / 0.8444 \\
			& & AWSRN-S                & 0.48$\,$M                  & 48.6G                & 34.02 / 0.9240 & 30.09 / 0.8376 & 28.92 / 0.8009 & 27.57 / 0.8391 \\ 
			& & \textbf{DRSAN-32s}     & \textbf{0.41$\,$M}         & \textbf{43.2G}      & \textbf{\textcolor{red}{34.41} / \textcolor{blue}{0.9272}} & \textcolor{red}{\textbf{30.27 / 0.8413}} & \textcolor{red}{\textbf{29.08 / 0.8056}} & \textcolor{red}{\textbf{28.19 / 0.8529}} \\ \cmidrule[0.1pt]{2-9}
			& \multirow{6}{*}{$<$ 1000$\,$K} & IDN                 & 0.55$\,$M                  & -      & 34.11 / 0.9253 & 29.99 / 0.8354 & 28.95 / 0.8013 & 27.42 / 0.8359 \\
			& & VDSR                 & 0.66$\,$M                  & 612.6G      & 33.66 / 0.9213 & 29.77 / 0.8314 & 28.82 / 0.7976 & 27.14 / 0.8279 \\
			& & MemNet                 & 0.68$\,$M                  & 2662.4G      & 34.09 / 0.9248 & 30.00 / 0.8385 & 28.96 / 0.8001 & 27.56 / 0.8376 \\
			& & IMDN                & 0.70$\,$M                  & -      & 34.36 / 0.9270 & 30.32 / 0.8417 & 29.09 / 0.8046 & 28.17 / 0.8519 \\
			& & \textbf{DRSAN-32m (Ours)} & \textbf{0.74$\,$M}         & \textbf{76.0G}       & \textcolor{red}{\textbf{34.50 / 0.9278}} & \textcolor{red}{\textbf{30.39 / 0.8437}} & \textcolor{red}{\textbf{29.13 / 0.8065}} & \textcolor{red}{\textbf{28.35 / 0.8566}} \\
			& & \textbf{DRSAN-48s (Ours)} & \textbf{0.75$\,$M}         & \textbf{78.0G}       & \textcolor{blue}{\textbf{34.47 / 0.9274}} & \textcolor{blue}{\textbf{30.35 / 0.8422}} & \textcolor{blue}{\textbf{29.11 / 0.8060}} & \textcolor{blue}{\textbf{28.26 / 0.8542}} \\ \cmidrule[0.1pt]{2-9}
			& \multirow{8}{*}{$<$ 2000$\,$K} & A$^2$F-M               & 1.00$\,$M                  & 100.0G               & 34.50 / 0.9278 & 30.39 / 0.8427 & 29.11 / 0.8054 & 28.28 / 0.8546 \\
			& & AWSRN-M                & 1.14$\,$M                  & 116.6G               & 34.42 / 0.9275 & 30.32 / 0.8419 & 29.13 / 0.8059 & 28.26 / 0.8545 \\
			& & SelNet                 & 1.16$\,$M                  & 120G                 & 34.27 / 0.9257 & 30.30 / 0.8399 & 28.97 / 0.8025 & -              \\
			& & A$^2$F-L               & 1.37$\,$M                  & 136.3G               & \textcolor{blue}{34.54 / 0.9283} & \textcolor{blue}{30.41 / 0.8436} & 29.14 / 0.8062 & 28.40 / 0.8574 \\
			& & AWSRN                  & 1.48$\,$M                  & 150.6G               & 34.52 / 0.9281 & 30.38 / 0.8426 & \textcolor{blue}{29.16 / 0.8069} & \textcolor{blue}{28.42 / 0.8580} \\
			& & CARN                   & 1.59$\,$M                  & 118.8G               & 34.29 / 0.9255 & 30.29 / 0.8407 & 29.06 / 0.8034 & 28.06 / 0.8493 \\
			& & DRCN                   & 1.77$\,$M                  & 17974G               & 33.82 / 0.9226 & 29.76 / 0.8311 & 28.80 / 0.7963 & 27.15 / 0.8276 \\ 
			& & \textbf{DRSAN-48m (Ours)}  & \textbf{1.29$\,$M}         & \textbf{133.4G}      & \textcolor{red}{\textbf{34.59 / 0.9286}} & \textcolor{red}{\textbf{30.42 / 0.8443}} & \textcolor{red}{\textbf{29.18 / 0.8079}} & \textcolor{red}{\textbf{28.52 / 0.8593}} \\ \midrule
			\multirow{20}{*}{$\times 4$} & \multirow{4}{*}{$<$ 500$\,$K} & FSRCNN                 & 0.01$\,$M                  & 4.6G                 & 30.48 / 0.8628 & 27.49 / 0.7503 & 26.90 / 0.7101 & 24.52 / 0.7221 \\
			& & DRRN                   & 0.30$\,$M                  & 6797G                & 31.68 / 0.8888 & 28.21 / 0.7720 & 27.38 / 0.7284 & 25.44 / 0.7638 \\
			& & AWSRN-SD               & 0.44$\,$M                  & 25.4G                & \textcolor{blue}{31.98 / 0.8921} & \textcolor{blue}{28.46 / 0.7786} & \textcolor{blue}{27.48} / \textcolor{red}{0.7368} & \textcolor{blue}{25.74 / 0.7746} \\ 
			& & \textbf{DRSAN-32s (Ours)}  & \textbf{0.41$\,$M}                  & \textbf{30.5G}                & \textcolor{red}{\textbf{32.15 / 0.8935}} & \textcolor{red}{\textbf{28.54 / 0.7813}} & \textbf{\textcolor{red}{27.54} / \textcolor{blue}{0.7364}} & \textcolor{red}{\textbf{26.06 / 0.7858}} \\ \cmidrule[0.1pt]{2-9}
			& \multirow{8}{*}{$<$ 1000$\,$K} & IDN                & 0.55$\,$M                  & -              & 31.82 / 0.8903 & 28.25 / 0.7730 & 27.41 / 0.7297 & 25.41 / 0.7632 \\
			& & AWSRN-S                & 0.59$\,$M                  & 37.7G                & 31.77 / 0.8893 & 28.35 / 0.7761 & 27.41 / 0.7304 & 25.56 / 0.7678 \\
			& & VDSR                & 0.66$\,$M                  & 612.6G                & 31.35 / 0.8838 & 28.01 / 0.7674 & 27.29 / 0.7251 & 25.18 / 0.7524 \\
			& & MemNet                 & 0.68$\,$M                  & 2662.4G              & 31.74 / 0.8893 & 28.26 / 0.7723 & 27.40 / 0.7281 & 25.50 / 0.7630 \\
			& & IMDN                & 0.72$\,$M                  & -                & 32.21 / \textcolor{blue}{0.8948} & \textcolor{blue}{28.58} / 0.7811 & 27.56 / 0.7353 & 26.04 / 0.7838 \\
			& & LapSRN                 & 0.81$\,$M                  & 149.4G              & 31.54 / 0.8850 & 28.19 / 0.7720 & 27.32 / 0.7280 & 25.21 / 0.7560 \\ 
			& & \textbf{DRSAN-32m (Ours)}  & \textbf{0.73$\,$M}                  & \textbf{49.0G}                & \textcolor{red}{\textbf{32.30 / 0.8954}} & \textcolor{red}{\textbf{28.66 / 0.7838}} & \textcolor{red}{\textbf{27.61 / 0.7381}} & \textcolor{red}{\textbf{26.26 / 0.7920}} \\
			& & \textbf{DRSAN-48s (Ours)}  & \textbf{0.73$\,$M}                  & \textbf{57.6G}                & \textbf{\textcolor{blue}{32.25} / 0.8945} & \textbf{28.55 / \textcolor{blue}{0.7817}} & \textcolor{blue}{\textbf{27.59 / 0.7374}} & \textcolor{blue}{\textbf{26.14 / 0.7875}} \\ \cmidrule[0.1pt]{2-9}
			& \multirow{8}{*}{$<$ 2000$\,$K} & A$^2$F-M               & 1.01$\,$M                  & 56.7G                & 32.28 / 0.8955 & 28.62 / 0.7828 & 27.58 / 0.7364 & 26.17 / 0.7892 \\
			& & AWSRN-M                & 1.25$\,$M                  & 72G                  & 32.21 / 0.8954 & 28.65 / 0.7832 & 27.60 / 0.7368 & 26.15 / 0.7884 \\
			& & A$^2$F-L               & 1.37$\,$M                  & 77.2G                & \textcolor{blue}{32.32} / \textcolor{red}{0.8964} & \textcolor{blue}{28.67} / 0.7839 & 27.62 / 0.7379 & \textcolor{blue}{26.32 / 0.7931} \\
			& & SelNet                 & 1.42$\,$M                  & 83.1G                & 32.00 / 0.8931 & 28.49 / 0.7783 & 27.44 / 0.7325 & -              \\
			& & AWSRN                  & 1.59$\,$M                  & 91.1G                & 32.27 / \textcolor{blue}{0.8960} & \textcolor{red}{28.69 / 0.7843} & \textcolor{red}{27.64} / \textcolor{blue}{0.7385} & 26.29 / 0.7930 \\
			& & CARN                   & 1.59$\,$M                  & 90.9G                & 32.13 / 0.8937 & 28.60 / 0.7806 & 27.58 / 0.7349 & 26.07 / 0.7837 \\
			& & DRCN                   & 1.77$\,$M                  & 17974G               & 31.53 / 0.8854 & 28.02 / 0.7670 & 27.23 / 0.7233 & 25.14 / 0.7510 \\
			& & \textbf{DRSAN-48m (Ours)}  & \textbf{1.27$\,$M}         & \textbf{88.7G}       & \textbf{\textcolor{red}{32.34} / \textcolor{blue}{0.8960}} & \textbf{28.65 / \textcolor{blue}{0.7841}} & \textbf{\textcolor{blue}{27.63} / \textcolor{red}{0.7390}} & \textbf{\textcolor{red}{26.33 / 0.7936}}       \\  \bottomrule[1pt]
		\end{tabular}
		}
	\end{center}
\end{table*}

\begin{table}[t]
	\caption{\label{heavysotatable} Quantitative comparisons of our model with the SOTA SR methods on benchmark datasets ($\times$4).}
	\renewcommand*{\arraystretch}{1.4}
	\begin{center}	
		\begin{tabular}{lrlllll}
			\hline
			Model   & Param   & Set5   & Set14 & BSDS100       & Urban100   \\ \hline
			EDSR    & 43.1$\,$M   & 32.46$\,$dB & 28.80$\,$dB & 27.71$\,$dB   & 26.64$\,$dB       \\ 
			RDN     & 22.3$\,$M   & 32.47$\,$dB & 28.81$\,$dB & 27.72$\,$dB   & 26.61$\,$dB     \\ 
			RCAN    & 15.6$\,$M   & 32.63$\,$dB & 28.87$\,$dB & 27.77$\,$dB   & 26.82$\,$dB     \\
			SAN     & 15.9$\,$M   & 32.64$\,$dB & 28.92$\,$dB & 27.78$\,$dB   & 26.79$\,$dB    \\
			DRN     & 9.8$\,$M    & \textbf{32.74$\,$dB} & \textbf{28.98$\,$dB} & \textbf{27.83$\,$dB}   & \textbf{27.03$\,$dB}    \\ \hline
			DRSAN   & \textbf{0.7$\,$M}    & 32.30$\,$dB & 28.66$\,$dB & 27.61$\,$dB   & 26.26$\,$dB     \\  \hline
		\end{tabular}
	\end{center}
\end{table}


\begin{table}[t]
	\caption{\label{timememorytable} Comparison of parameter, memory usage, inference time and performance between SR methods. Memory represents GPU memory used by total tensor of each model. Inference time is calculated on B100 ($\times$4) dataset}. 
	\renewcommand*{\arraystretch}{1.4}
	\begin{center}	
		\begin{tabular}{lrrrll}
			\hline
			Model       & Param       & Memory         & Running time     & PSNR           \\ \hline
			CARN-M      & 0.41$\,$M       & \textbf{3.0MB}          & 10.473ms        & 27.44$\,$dB        \\ 
			VDSR        & 0.66$\,$M       & 3.7MB          & 22.618ms        & 27.29$\,$dB        \\ 
			A$^2$F-L    & 1.37$\,$M       & 11.7MB         & 16.524ms        & 27.62$\,$dB        \\ 
			DRN-S       & 4.8$\,$M        & 20.7MB         & 41.519ms        & \textbf{27.78$\,$dB}        \\ 
			RCAN        & 15.6$\,$M       & 61.3MB         & 75.448ms        & 27.76$\,$dB        \\ \hline
			DRSAN-32s   & \textbf{0.41$\,$M}       & 3.4MB          & \textbf{8.291ms}        & 27.54$\,$dB        \\
			DRSAN-48m   & 1.27$\,$M       & 6.7MB          & 13.317ms        & 27.63$\,$dB        \\ \hline
		\end{tabular}
	\end{center}
\end{table}

\subsection{Model Analysis}
\begin{figure}[t]
	\begin{center}
		\includegraphics[width=\linewidth]{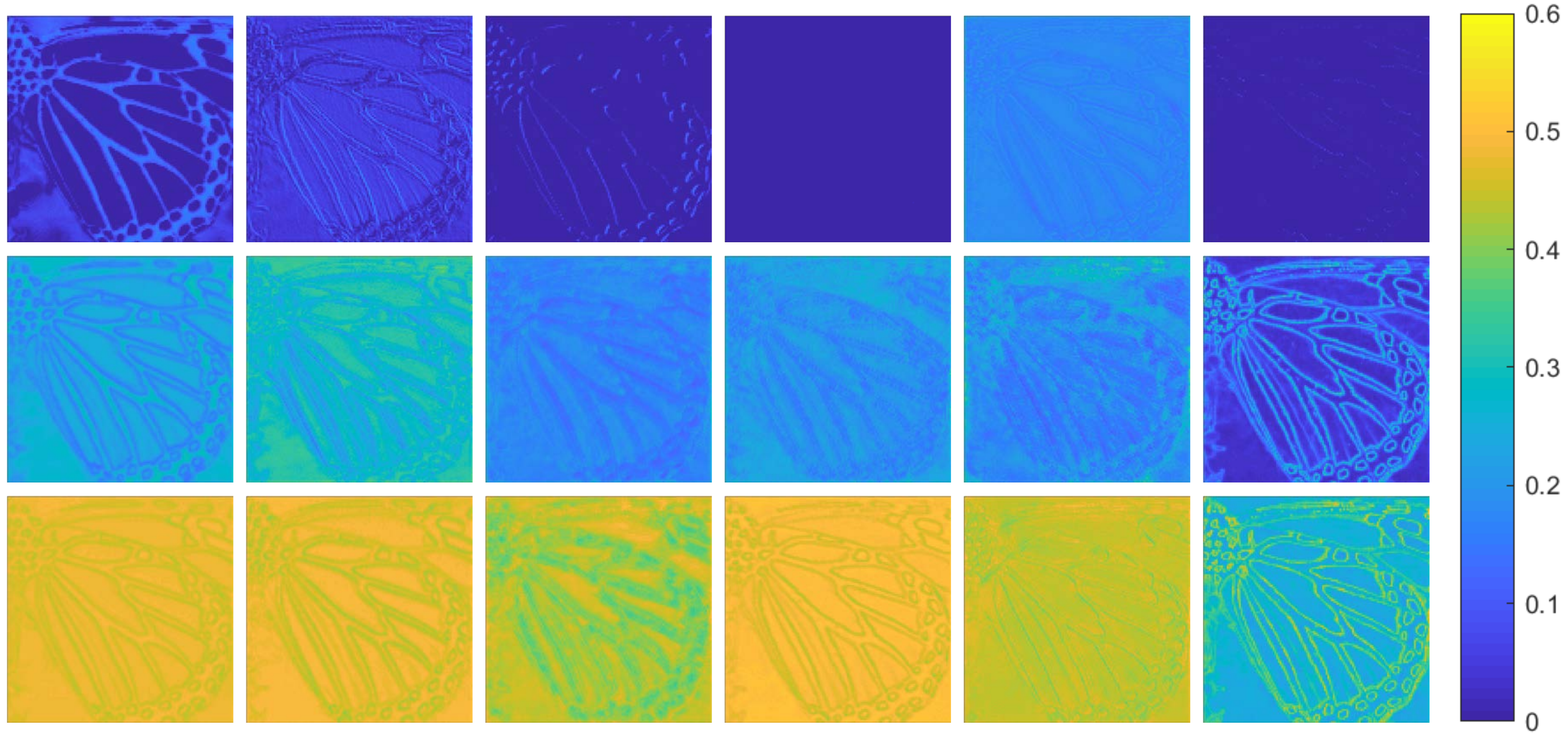}
	\end{center}
	\caption{Activation maps of the SA (the first row), PA (second), and DRSA (third) from the residual blocks of the networks. For visualization, the results of PA and DRSA are obtained by averaging channel-wise attention maps.}
	\label{fig:visualizedactivationmap}
\end{figure}
\begin{figure}[!ht]
	\begin{center}
		\includegraphics[width=\linewidth]{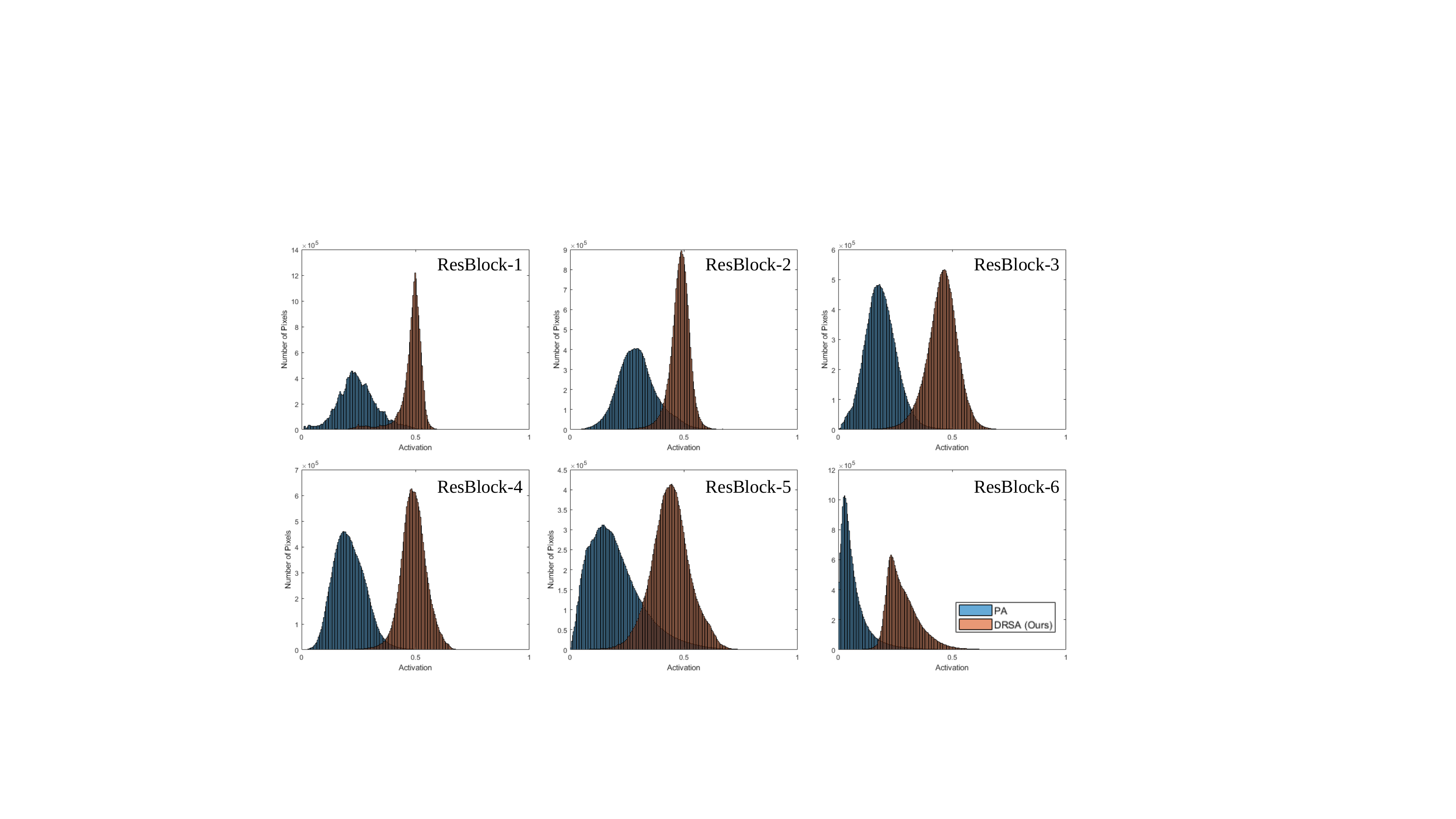}
	\end{center}
	\caption{Histogram of activation values of PA and DRSA. Each histogram represents the frequency distribution of the activation values from each block of the network. Activation values are collected from images of the Set5 ($\times2$) dataset.}
	\label{fig:histograms}
\end{figure}
In this section, we perform additional experiments to understand how DRSA behaves in the network and its differences compared with previous works.
\subsubsection{Activation Analysis}
First, we compare the activation maps with the networks using SA, PA, and DRSA. The activation maps are shown in Fig. {\ref{fig:visualizedactivationmap}}, obtained by averaging attention maps in a channel. Unlike PA and DRSA, SA produces a 2D attention map (H$\times$W$\times$1) and applies the same attention weights to all channels of the feature. Each channel in the feature maps may require different attention maps for the optimal solution, and this lack of flexibility leads to almost inactivated attention maps, such as the first row's 4th and 6th attention maps of SA, implying that the 2D attention scheme is less helpful for some residual blocks.

However, as shown in the 2nd and 3rd columns of Fig. {\ref{fig:visualizedactivationmap}}, the attention schemes yielding 3D attention maps (H$\times$W$\times$C) are always active in any residual blocks, implying that each channel in the residual feature may require different attention weights for the optimal results. Additionally, note that there is some shared tendency between PA and DRSA in attention weight decision policy. We observe that both attention methods attempt to assign higher weights to the texture part, except for the last residual block. In contrast, the edge part has higher attention weights in the last residual block located immediately before the upsampler network. This result indicates that both attention schemes work differently depending on the location, especially at the network's last residual block.

Next, we compare the histogram of activation maps with two different attention models to explore the influence of the proposed DRSA. Each histogram in Fig. {\ref{fig:histograms}} represents the activation map at each network's residual block. The histograms in Fig. {\ref{fig:histograms}} are obtained by using Set5 ($\times2$) images as the input. As shown in Fig. {\ref{fig:histograms}}, PA has values that range near zero, making the network suppress the residual block's feature values. On the other hand, the DRSA has values that range near 0.5, enabling the network to scale the features with higher values than those of the PA. In other words, our DRSA tends to boost more attentive features and suppress less informative features.

Therefore, the PA mostly turns off the unimportant feature values for the last residual block, when struggling with the limited range near 0. In contrast, our DRSA rescales the feature values with more discriminative importance, assigning different attention values even with the less important features.
\begin{figure}[t]
	\begin{center}
		\includegraphics[width=\linewidth]{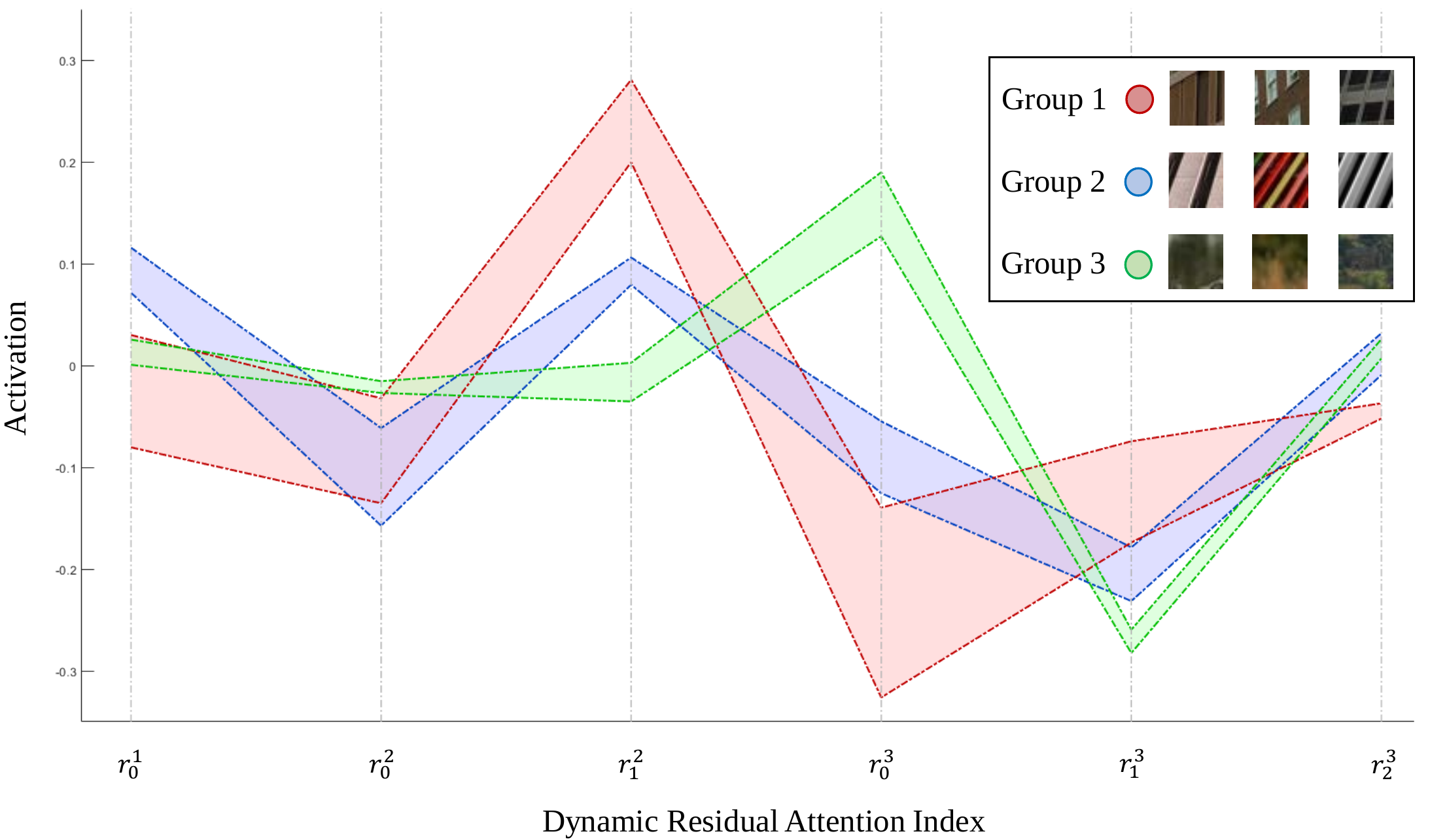}
	\end{center}
	\caption{The activation values of DRA in the 1st DRAG using different patches as input. Patches with similar DRA values are grouped. Patches are collected from images of benchmark datasets ($\times$2).}
	\label{fig:activemap}
\end{figure}

\subsubsection{Behavior Analysis}
As stated previously, we devised DRA to have different values by the DRM, depending on input image properties.
To investigate this, we group input patches with similar properties and compare their DRA value distributions in Fig.~{\ref{fig:activemap}}.
We use DRA values of the first DRAG of the ablation model used in the ablation study for this visualization. We can see that the variations of DRA values have different forms depending on image groups. Specifically, the images in Group 1 have rectangular textures, Group 2 has strong edges, Group 3 has almost no edges or textures, and DRA values have different behavior depending on the group. Notably, we observe that the variation of DRA values of Group 3, which lacks edges and high-frequency information, is relatively small. On the other hand, the DRA values dynamically change depending on the shapes of edges and textures. We may interpret that the edge and high-frequency textures are discriminative features for our DRA, as their values change dynamically to find effective features for reconstructing the final result.

\begin{figure}[t]
	\begin{center}
		\includegraphics[width=\linewidth]{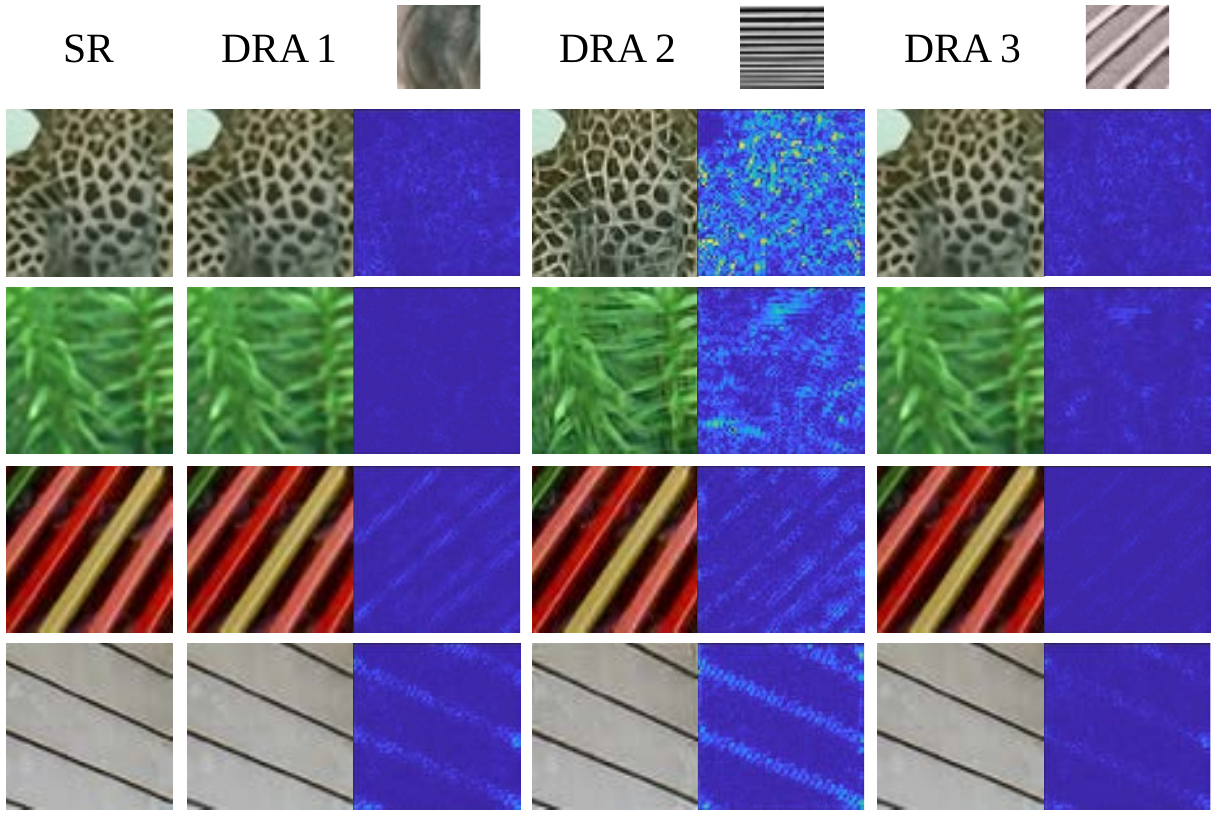}
	\end{center}
	\caption{The reconstructed images using DRA from different patches and their visualized difference maps. The difference map is calculated on the Y channel of the image and its original SR image. Patches are collected from images of benchmark datasets ($\times$2).}
	\label{fig:ablationdra}
\end{figure}

To examine whether the DRA values affect the SR reconstruction, we reconstruct specific images using DRAs extracted from different images with different characteristics, as shown in Fig.~{\ref{fig:ablationdra}}. The first "SR" column shows the images to be super-resolved, and "DRA$n$" columns are the super-resolved images when the DRA values are given by each of the image patches shown in the first row. Specifically, DRA1 is the case in which DRA values are given by the smooth patch, DRA2 by the strong horizontal edges, and DRA3 by diagonal edges.
Additionally, the blue figure columns on the right side of DRA$n$ columns are the differences from the corresponding original SR results. 
The figure shows that the changes in DRA lead to changes in the SR results of the network. For example, the 4th and last rows show that the stripe patterns are best reconstructed by the DRA learned from similar patches, as the differences are small for this case (last column). Additionally, most of the changes are observed near edges in images, suggesting that DRA is mainly involved in reconstructing the edge portion of the image. In summary, the results imply that DRA helps SR reconstruction by focusing on edge-related features. We hope that our study may provide a small clue to understanding the features of SR networks for future research.

\begin{figure*}[t]
	\begin{center}
		\includegraphics[width=0.75\linewidth]{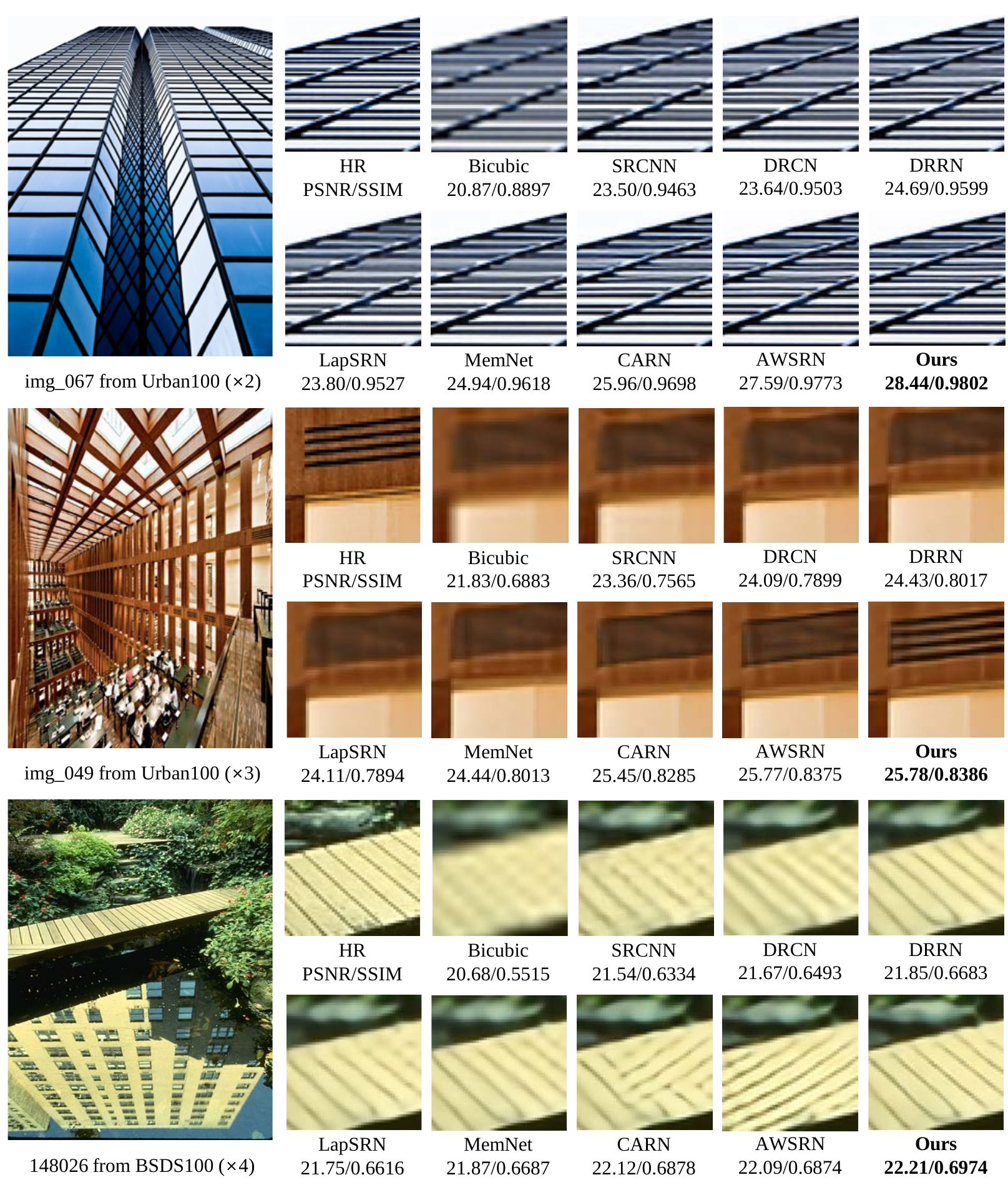}
	\end{center}
	\caption{Qualitative comparisons with the bicubic degradation model on the BSDS100 and Urban100 datasets. The best results are marked in \textbf{bold}.}
	\label{fig:visualizedresults}
\end{figure*}

\subsection{Comparison with state-of-the-art methods}
In this section, we compare the proposed DRSAN with several lightweight SR networks and networks with parameters similar to lightweight methods, including FSRCNN \cite{dong2016accelerating}, VDSR \cite{Kim_2016_VDSR}, DRCN \cite{kim2016deeply}, DRRN \cite{tai2017image}, MemNet \cite{tai2017memnet}, SelNet \cite{choi2017deep}, IDN \cite{hui2018fast}, IMDN \cite{hui2019lightweight}, LapSRN \cite{lai2018fast}, CARN \cite{ahn2018fast}, FALSR \cite{chu2019fast}, OISR \cite{he2019ode}, AWSRN \cite{wang2019lightweight}, and A$^2$FNet \cite{wang2020lightweight}.
Additionally, we compare our method with the heavyweight SOTA SR methods requiring more than $2\,$M parameters, including EDSR \cite{lim2017enhanced}, RDN \cite{zhang2018residual}, RCAN \cite{zhang2018image}, SAN \cite{dai2019second}, and DRN \cite{guo2020closed}.
For comparing the complexities of the networks, we compare the number of parameters and the number of operations by Multi-Adds, which is the number of multiply-accumulate operations for a single 720p (1280$\times$720) HR image. For detailed comparisons between SR methods with various network capacities, we split the comparison results into several groups, according to the upscaling factor and the number of parameters.

Table \ref{sotatable} shows the comparison of complexity and performance between lightweight SR models. Our models perform better than existing methods in the same parameter scope, indicating an efficient trade-off between reconstruction performance and complexity. Note that our DRSAN-32m model shows comparable performance to the AWSRN-M and A$^2$F-M models, which have more parameters and operations than ours. Additionally, we observe that DRSAN-32 models outperform DRSAN-48 models with similar network capacities. It suggests increasing the depth, rather than the channels, to obtain a better performance with similar network complexity.

Table \ref{heavysotatable} compares the complexity and performance between the heavyweight SOTA SR methods and our DRSAN-32m. It can be seen that our model requires at least ten times fewer parameters than the heavy models while delivering competitive results. Precisely, our lightweight network shows a similar performance ($0.1 \sim 0.2\,$dB) as heavy networks in the case of the BSD100 dataset, and the difference becomes larger ($0.4 \sim 0.8\,$dB) in the case of the Urban100 dataset, which consists of images having complex structures.

In Table \ref{timememorytable}, we compare the inference time and memory usage between the SOTA methods and ours. Compared with lightweight models with similar parameters, our methods show remarkable performances while requiring less inference time. Compared with heavy SR methods, our methods require three to nine times less memory and inference time while sacrificing $0.1$ to $0.2\,$dB of performance. These observations suggest that our method has an efficient trade-off between model complexity and performance.

Finally, we show qualitative comparisons on the BSDS100 and Urban100 datasets in Fig. \ref{fig:visualizedresults}. We can see that other methods struggle with blurring and aliasing artifacts at the challenging stripe patterns. In contrast, our method reconstructs more details and reduces the blurring artifacts. These results indicate that our dynamic residual network scheme shows a comparable performance to handcrafted networks by using the proper combination of residual features considering the input image characteristics.


\subsection{Ablation Study}
\begin{table}[t]
	\caption{\label{ablationdrsa} Ablation study on the proposed DRSA. Res is the original residual connection from SRResNet, and All Res indicates all possible residual connections in DRAG.} 
	\renewcommand*{\arraystretch}{1.4}
	\begin{tabular}{|c|c|c|c|c|l|l|}
		\hline
		\multirow{2}{*}{Res} & \multirow{2}{*}{All Res} & \multicolumn{3}{c|}{DRAG}                  & \multirow{2}{*}{Params} & \multirow{2}{*}{PSNR} \\ \cline{3-5}
		&                                   & DRA          & RSA          & Concat       &                         &                       \\ \hline
		$\checkmark$         &                                   &              &              &              & 357$\,$K                    & 31.54$\,$dB               \\
		$\checkmark$         &                                   &              & $\checkmark$ &              & 357$\,$K                    & 31.66$\,$dB               \\
		$\checkmark$         &                                   &              & $\checkmark$ & $\checkmark$ & 375$\,$K                    & 31.73$\,$dB               \\
		&                                   & $\checkmark$ & $\checkmark$ &              & 358$\,$K                    & 31.71$\,$dB               \\
		& $\checkmark$                      &              & $\checkmark$ & $\checkmark$ & 375$\,$K                    & 31.68$\,$dB               \\
		&                                   & $\checkmark$ &              & $\checkmark$ & 377$\,$K                    & 31.70$\,$dB               \\
		&                                   & $\checkmark$ & $\checkmark$ & $\checkmark$ & 377$\,$K                    & \textbf{31.82$\,$dB}               \\ \hline
	\end{tabular}
\end{table}
\begin{table}[t]
	\caption{\label{ablationactivation} Comparison of the network performance between sigmoid and hyperbolic tangent (Tanh) as an activation function of DRM.} 
	\renewcommand*{\arraystretch}{1.4}
	\begin{center}	
		\begin{tabular}{llll}
			\hline
			Activation   & Sigmoid   & Tanh      & None  \\ \hline
			PSNR   & $31.77\,$dB   & 31.79$\,$dB   & \textbf{31.82$\,$dB}     \\ \hline
		\end{tabular}
	\end{center}
\end{table}
\begin{table}[!ht]
	\caption{\label{ablationattention} Comparison of channel attention (CA), spatial attention (SA), pixel attention (PA), and the proposed DRSA. The results show average PSNR evaluated on the Set5, Set14, BSDS100, and Urban100.} 
	\renewcommand*{\arraystretch}{1.4}
	\begin{center}	
		\begin{tabular}{llllll}
			\hline
			Model  & Baseline   & CA   & SA      & PA      & DRSA (Ours) \\ \hline
			Params & 375$\,$K       & 378$\,$K & 376$\,$K    & 389$\,$K    & 377$\,$K        \\
			PSNR   & 32.06$\,$dB    & 32.13$\,$dB & 32.10$\,$dB & 32.16$\,$dB & \textbf{32.18$\,$dB}     \\ \hline
		\end{tabular}
	\end{center}
\end{table}

\subsubsection{Dynamic Residual Connection}

We first demonstrate the effectiveness of the DRAG structure by replacing its components with well-known methods. To examine the influence of the DRA, we replace it with a residual connection from SRResNet \cite{ledig2017photo} and all possible residual connections, which amounts to {\em DRA without the parameters} $r$. Additionally, we examine the effect of the RSA by applying it to two different residual connection methods: residual connection from SRResNet and our dynamic residual connection. As a baseline network, we use the DRSAN structure with 2 DRAGs trained with the DIV2K dataset.

Table {\ref{ablationdrsa}} shows the performance of ablation models for Urban100 ($\times2$) with different configurations. The first and second rows show that the RSA mechanism boosts $0.12\,$dB without any parameter increase. This tendency can also be seen in the 6th and 7th rows, which supports that the proposed RSA improves the network performance by jointly working with residual connections. The 3rd, 5th, and 7th rows indicate that simply connecting all preceding residual features does not improve the performance and sometimes even deteriorates its performance. These results show that the DRM properly recalibrates the weights of residual features.

Through extensive experiments, it was found that obviating the sigmoid activation function gives a better performance than using the sigmoid activation function in our case. However, the sigmoid generally helps stable convergence and obtains a better performance in most cases. We conjecture that the sigmoid can hamper the diversity of feature combinations necessary for our scheme because the sigmoid limits the range of output features between 0 and 1.
The OISR, which is an SR network inspired by the ordinary differential equation \cite{he2019ode}, also scales the residual features to a value of 2 or $-1$, suggesting that it is more advantageous not to limit the range of scaling values. Table \ref{ablationactivation} compares the performance of ablation models with different activation functions to see the change in performance depending on the activation function. As shown in Table \ref{ablationactivation}, using hyperbolic tangent rather than sigmoid proves to perform slightly better. Furthermore, DRM without the activation function achieves better results than DRM with a hyperbolic tangent, indicating that limiting the range of scaling values degrades the network performance.


\subsubsection{Residual Self-Attention}
To examine the effectiveness of the proposed attention scheme, we conduct ablation experiments by replacing DRSA with popular attention schemes in SR, namely CA, SA, and PA. We use the same configurations in \cite{woo2018cbam} for the CA and the SA. As a baseline, we use the DRSAN structure with two residual groups but without the DRSA.

The results are shown in Table \ref{ablationattention}, which compares the performance of attention schemes for Set5, Set14, BSDS100, and Urban100 ($\times2$). As shown in Table \ref{ablationattention}, CA and SA improved the PSNR by $0.07\,$dB and $0.04\,$dB, respectively, with small increases of $1\,$K to $3\,$K in the number of parameters. Unlike 1D CA and 2D SA, PA \cite{zhao2020efficient} produces a 3D attention map with pointwise convolution (1$\times$1) for an accurate SR reconstruction. The PA brings a PSNR gain of $0.10\,$dB with a $13\,$K parameter increase. However, our DRSA shows a $0.12\,$dB performance boost with the rise of $2\,$K parameters, indicating that our attention scheme is competitive and efficient compared with previous works.

\section{Conclusion}
We have proposed a dynamic residual network scheme for a lightweight SR system, utilizing diverse combinations of residual features considering the input statistics. Additionally, we have introduced RSA that boosts the network performance without additional modules by cooperating with the residual structure. The network design schemes and attention mechanisms are easily applicable for other residual networks without designing a complex network structure. Experimental results on well-known benchmark datasets have shown that our method achieves better or comparable performances to SOTA models with complex configurations. We will extend the dynamic residual network scheme to other restoration tasks as future work and release our code at https://github.com/saturnian77/DRSAN.

%
%
%

\ifCLASSOPTIONcaptionsoff
  \newpage
\fi

\begin{IEEEbiographynophoto}{Karam Park}
received the BS degree in Electrical and Computer Engineering from Seoul National University, Seoul, Korea, in 2018. He is currently studying for a Ph.D. degree in Electrical and Computer Engineering from Seoul National University, Seoul, Korea. His main research interest is image processing based on deep learning.
\end{IEEEbiographynophoto}

\begin{IEEEbiographynophoto}{Jae Woong Soh}
received the BS degree in Electrical and Computer Engineering from Seoul National University, Seoul, Korea, in 2016. He is currently studying for a Ph.D. degree in Electrical and Computer Engineering from Seoul National University, Seoul, Korea. His research interests include image processing, computer vision, and machine learning.
\end{IEEEbiographynophoto}

\begin{IEEEbiographynophoto}{Nam Ik Cho}
received the BS, MS, and Ph.D. degrees in Control and Instrumentation Engineering from Seoul National University, Seoul, Korea, in 1986, 1988, and 1992. From 1991 to 1993, he was a research associate with the Engineering Research Center for Advanced Control and Instrumentation, Seoul National University. From 1994 to 1998, he was with the University of Seoul as an assistant professor of Electrical Engineering. In 1999, he joined the Department of Electrical and Computer Engineering, Seoul National University, where he is currently a professor. His research interests include image processing, adaptive filtering, digital filter design, and computer vision.
\end{IEEEbiographynophoto}



\end{document}